\documentclass{raa}            

\usepackage{graphicx,times}             
\newcommand{\ldri}{LAMOST DR1}

\begin{document}

   \title{Searching for Classical Be Stars from LAMOST DR1}

   \volnopage{Vol.0 (200x) No.0, 000--000} 
   \setcounter{page}{1}     
   \author{Lin, C.~C. 
      \inst{1}
   \and Hou, J.~L. 
      \inst{1}
   \and Chen, L. 
      \inst{1}
   \and Shao, Z.~Y. 
      \inst{1}
   \and Zhong, J. 
      \inst{1}
   \and Yu, P.~C. 
      \inst{2}
   }

   \institute{Shanghai Astronomical Observatory, Chinese Academy of Sciences,
             80 Nandan Road, Shanghai 200030, China; {\it cclin@shao.ac.cn}\\
        \and
             Department of Physics and Astronomy, University of California,
             Los Angeles, CA 90024, USA\\
   }

   \date{Received~~2009 month day; accepted~~2009~~month day}
\abstract{
We report on searching for Classical B-type emission-line (CBe) stars 
from the first data release (DR1) of the Large Sky Area Multi-Object fiber 
Spectroscopic Telescope (LAMOST; also named the Guoshoujing Telescope). 
A total of 192 (12 known CBes) objects were identified as CBe candidates with 
prominent He~I~$\lambda4387$, He~I~$\lambda4471$, and Mg~II~$\lambda4481$ 
absorption lines, as well as H$\beta$~$\lambda4861$ and H$\alpha$~$\lambda6563$ 
emission lines.  These candidates significantly increases current CBe sample of 
about 8\%.  Most of the CBe candidates are distributed at the Galactic 
Anti-Center due to the LAMOST observing strategy.  Only two of CBes are in the 
star clusters with ages of 15.8 and 398~Myr, respectively.
\keywords{stars: emission-line, Be, stars: early-type, 
 open clusters and associations: general}
}

   \authorrunning{Lin et al.}            
   \titlerunning{LAMOST Classical Be Stars}  

   \maketitle
\section{Introduction}          
Classical Be stars (CBes) are non-supergiants B-type stars symbolized by Balmer 
series, mostly emit at H$\alpha$~$\lambda6563$ and H$\beta$~$\lambda4861$ 
lines.  Apart from the emission phenomenon, CBes have fast rotation with an 
equatorial speed up to $70-80\%$ of breakup velocity 
(Porter \& Rivinius~\cite{por03}).  CBes exist not only in young or 
intermediate age star clusters but also isolated in the field.  In the past, 
the astronomers concentrated on analysis of the properties of bright CBe with 
their individual observational data.  Nowadays, large sky photometric surveys 
have made it possible to study homogeneously and completely of CBes, e.g., 
Zhang et al.~(\cite{zha05}) analyzed the infrared color of 1185 CBes stars with 
the 2MASS (Cutri et al.~\cite{cut03}) data.  Raddi et al.~(\cite{rad15}) 
present a catalogue of 247 photometrically and spectroscopically confirmed CBes 
in the direction of the Perseus Arm of the Milky Way from the IPHAS 
(Drew et al.~\cite{dre05}).  Furthermore, Chojnowski et al.~(\cite{cho15}) 
discovered 128 new CBes and increased the total number of known CBe sample by 
$\sim6\%$ from SDSS-III/APOGEE (Eisenstein et al.~\cite{eis11},  
Majewski~\cite{maj12}).  

Recently, the Be Star Spectra database (BeSS\footnote{http://basebe.obspm.fr}) 
with no more than 3000 CBes was created by Neiner et al.~(\cite{nei11}), in 
order to collect all existing and future Be star spectra for the statistical 
studies in the Be star community.  However, the sample of CBes remains 
inhomogeneous because the wide-field spectroscopic observations are time 
consuming and often limited to bright stars.  On the other hand, the sample of 
CBes is incomprehensive due to the insufficient information of ages or 
distances.  With a large field of view and the highest spectral acquisition 
rate, the Large Sky Area Multi-Object Spectroscopic Telescope (LAMOST) survey 
thus provides us an excellent opportunity to conduct a systematic survey for CBes.  

We report on a search of CBes from the LAMOST First Data Release, hereafter 
\ldri.  We have developed an algorithm to identify CBes, and visually 
inspected their spectra for confirmation.  In Section 2, we described the 
acquisition of the observations and the methodology of recognizing H$\alpha$ 
emission stars.  In Section 3, we discuss the results and give a summary of 
this study in section 4.

\section{The Data and Searching Methodology} 
The major dataset used for this study was from the \ldri.  Additionally, the 
2MASS point source catalog has also been used to supplement the photometric 
analysis.  

\subsection{The \ldri}
The LAMOST\footnote{http://www.lamost.org/}, also named the Guoshoujing 
Telescope, is a quasi-meridian reflecting Schmidt telescope located at Xinglong 
Observing Station in the Hebei province of China.  The telescope has an 
effective aperture of 3.6--4.9~m, and a field of view of about $5^{\circ}$ in 
diameter.  A total of 16 low-resolution spectrographs, 32 CCDs, and 4000 
fibers are mounted on the telescope.  Each spectrograph has a spectral 
resolution of $R\sim1800$ in the wavelengths ranging from 3700~\AA\ to 
9000~\AA\ (Cui et al.~\cite{cui12}; Zhao et al.~\cite{zha12}). 

The \ldri\ includes more than two million spectra with a limiting magnitude of 
$r\sim18.5$~mag that are obtained from the pilot survey and first year general 
survey (Luo et al.~\cite{luo12}; Luo et al.~\cite{luo15}).  The \ldri\ also has 
stellar catalogs of about 1.2~million spectroscopically classified stars with 
their atmospheric parameters such as radial velocities, effective temperatures, 
surface gravities, and metallicities.  About a quarter of million stars are 
somehow un-classified or not well-classified, which might be due to the 
interstellar extinction at the short wavelength range resulting the fitting 
failure in the \ldri.  In order to identify a large sample of CBe candidates, 
we analyzed the whole \ldri\ dataset with the mean signal-to-noise ratio 
$SNR\ge10$.  

\subsection{The Search Algorithm}
The major indicator of a B-type star is the set of hydrogen Balmer absorption 
lines and in conjunction with some neutral helium (He~I) absorption lines or an 
ionized magnesium (Mg~II) absorption line.  CBes in particular feature hydrogen 
emission lines mostly at H$\alpha$ and H$\beta$, but fade through the rest of 
the Balmer series.  Therefore, we focus on searching for stars with prominent 
He~I~$\lambda4387$, He~I~$\lambda4471$, and Mg~II~$\lambda4481$ absorption 
lines, as well as those with H$\beta$~$\lambda4861$ and H$\alpha$~$\lambda6563$ 
emission lines.  

To quantify these line indexes, we calculated the equivalent width 
($EW_{\lambda}$) of each line by the following equation   
\begin{equation}
EW_{\lambda} = \int 1-f_{l}/\bar{f_{c}}~d\lambda, 
\end{equation}
where $f_{l}$ is the flux of each line and $\bar{f_{c}}$ is the average of 
local pseudo-continuum estimated within 140~\AA\ width at each line.  The 
integration range covered a width of 10~\AA.  The empirical line 
$EW_{\lambda}$ of known CBes observed by the LAMOST are estimated and 
summarized in Table~\ref{ew}.  The CBe candidates are required 
to be satisfied with the similar $EW_{\lambda}$ of known CBes at 
He~I~$\lambda4387$, He~I~$\lambda4471$, and Mg~II~$\lambda4481$ lines and 
those with $EW_{\lambda}$ less than 0.33~\AA\ and 0.50~\AA\ at 
H$\beta$~$\lambda4861$ and H$\alpha$~$\lambda6563$ lines, respectively.

\begin{figure}
\centering
\includegraphics[width=\textwidth]{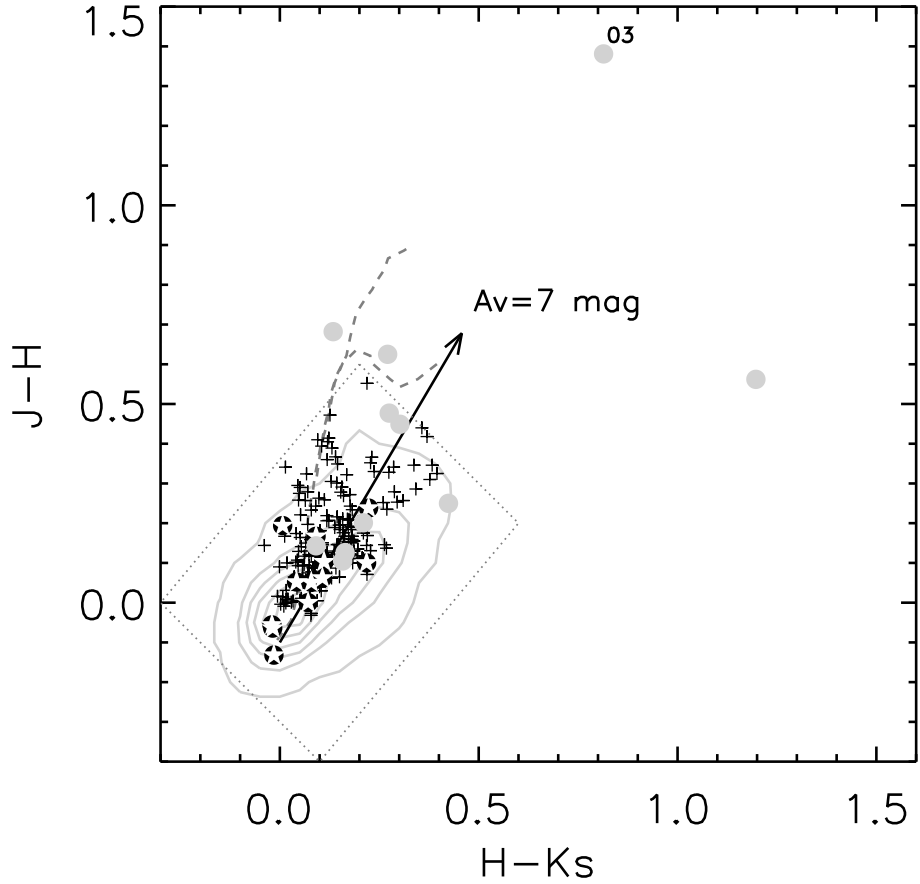}
\caption{The 2MASS color-color diagram.  The black dashed lines show the giant 
(upper) and dwarf (lower) loci (Bessell \& Brett~\cite{bes88}) converted to the 
2MASS system.  The arrow represents the reddening direction 
(Rieke \& Lebofsky~\cite{rie85}) for typical Galactic interstellar extinction 
(RV = 3.1).  The gray contours demonstrate known CBes distribution, and the 
dotted-box is defined as Be region to include most of CBes.  The black crosses 
are CBe candidates.  The known CBes are marked with white asterisks and grey 
filled circles for identified and unidentified, respectively.}
\label{tcd}
\end{figure}

\begin{figure}
\centering
\includegraphics[width=\textwidth]{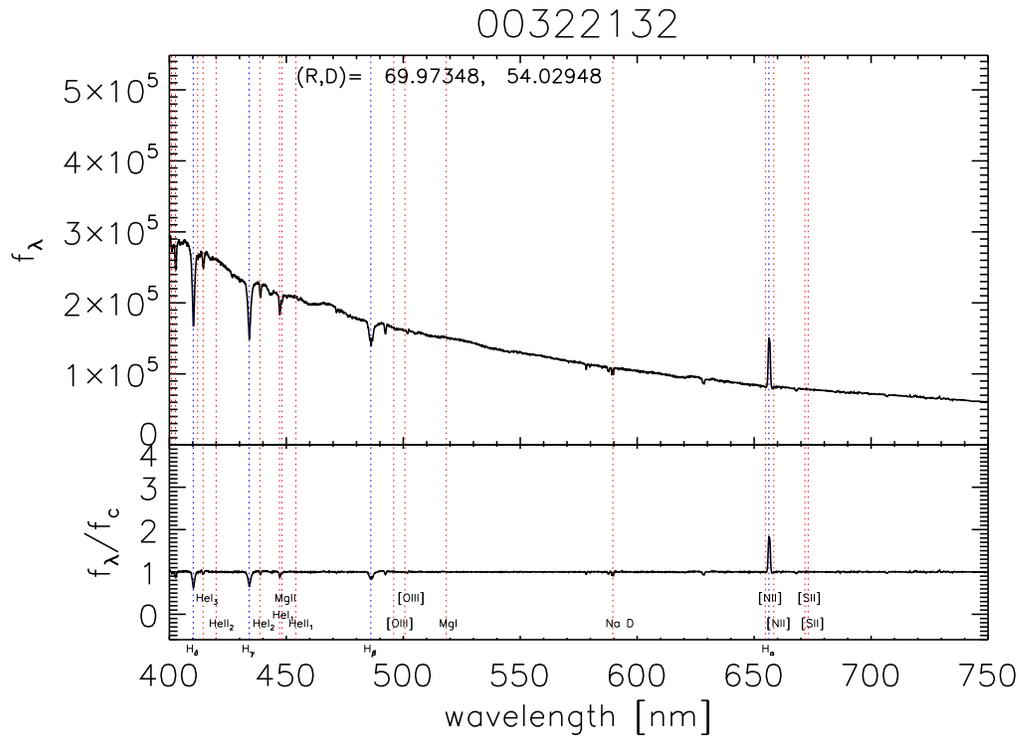}
\caption{A spectrum of one newly found CBe in \ldri.  The flux in upper panel 
is relative flux.  Blue dashed lines indicate the Balmer series, and red dashed 
lines represent some major lines.  The lower panel shows normalized spectra 
with respect to the pseudo-continuum.  Color version can be seen online.}
\label{spectra}
\end{figure}

To rule out contamination of B[e] or Herbig Ae/Be stars, the CBe 
candidates are also required to have the similar colors of known CBes.  We 
defined a ``Be region'' with $J-H$ versus $H-K_{s}$ color-color diagram.  
The Be region could cover most Be stars that were collected from the 
literatures (Zhang et al.~\cite{zha05}).  As shown in Figure~\ref{tcd}, 
gray contours represent over 1000 known CBes.  Assuming that most CBes have 
similar infrared colors, we thus could select CBe candidates inside the 
gray-dotted region in the color-color diagram.  From the \ldri, we finally 
identified 192 CBe candidates.  Among these candidates, 180 are newly 
discovered CBe candidates and 12 are known CBes.  Figure~\ref{spectra} 
demonstrates one example of the CBe candidates with the H$\alpha$ emission 
line, a very weak emission superposes on the H$\beta$ absorption line, and 
He~I~${\lambda4387}$ and He~I~${\lambda4471}$ absorption lines.

\section{Results and Discussion}
Although there are more than 3000 CBes that have been presented by previous 
studies (Neiner et al.~\cite{nei11}; Zhang et al.~\cite{zha05}; 
Raddi et al.~\cite{rad15}), only 23 CBes were cross-matched in the \ldri.  
Among these observed CBes, 12 ones are re-identified.  Some CBes 
are not re-identified due to low $SNR$ (5 stars) and poor calibration (4 stars).  
Another un-identified known CBe K03, also named GSC 02342-00359, is a young 
stellar object and classified to be an F0-G4 star in NGC 1333 with large 
reddening as seen in Figure~\ref{tcd} (Liu et al.~\cite{liu80}).  The reason 
may imply to the H$\alpha$ variability of CBes so that we probably could not 
identify its H$\alpha$ emission line phenomenon at certain epoch 
(Rivinius et al.~\cite{riv13}).  There is only one known CBe K10, missed to 
be re-identified because of the weak Mg~II absorption.  Therefore, excluding 
the spectra with poor calibration and low $SNR$, the detection rate of CBes is 
about 85\%.  The known CBes are listed with the 2MASS magnitudes in 
Table~\ref{know} and some bright ($J < 9$~mag) CBes with matched radii larger 
than 5\arcsec\ are listed below the star K13.  

\begin{figure}
\centering
\includegraphics[width=\textwidth]{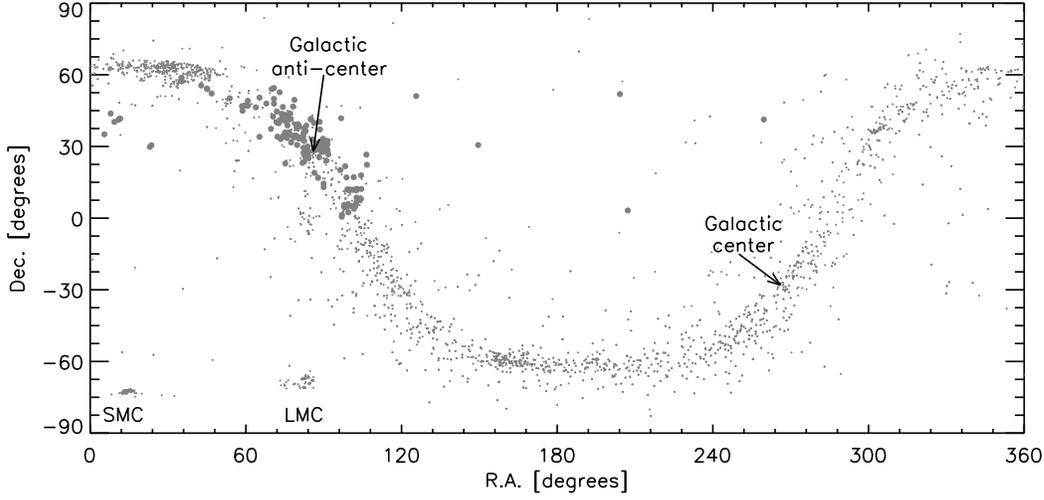}
\caption{The spatial distribution of CBes.  The small gray dots are known 
CBes from Zhang et al.~(\cite{zha05}), Neiner et al.~(\cite{nei11}), 
and Raddi et al.~(\cite{rad15}).  The large dots are the LAMOST CBe candidates.  
The Galactic Center/Anti-Center and Magellanic Clouds are marked.}
\label{spatial}
\end{figure}

The new CBe sample significantly increases about 8\% of current sample.  The 
CBe candidates are listed with the 2MASS magnitudes in Table~\ref{can} and the 
SIMBAD\footnote{http://simbad.u-strasbg.fr/simbad/} objects are noted in the 
last column.  Although some of these candidates have been identified as 
emission line stars by Kohoutek \& Wehmeyer~(\cite{knw97})\footnote{
http://www.hs.uni-hamburg.de/DE/Ins/Per/Kohoutek/index.html}, 
the spectral types were not yet confirmed until the LAMOST observations.  
The spatial distribution of CBe candidates from \ldri\ is shown in 
Figure~\ref{spatial}.  Most of CBe candidates were found along the Galactic 
plane that is similar to the trend seen in previous studies.  And CBe 
candidates are more concentrated toward the Galactic Anti-Center because of 
the observing strategy.

The distance and age of CBes can be determined if they are members of star 
clusters.  Using the method of membership identification presented by 
Yu et al.~(\cite{yup15}) that is based on photometric isochrone, spatial 
distribution, and proper motions, we found that only two CBes are the member 
of open clusters.  The CBe L032 is a member of open cluster Kronberger\,18 
with an age of $\sim$15.8~Myr and with a distance of 2700~pc 
(Kharchenko et al.~\cite{kha13}).  The other CBe L056 is a member of open 
cluster FSR\,1025 with an age of $\sim$398~Myr and with a distance of 2095~pc 
(Kharchenko et al.~\cite{kha13}).  The upper limit of the CBes age thus extends 
to 398~Myr which is older than the previous studies by 
McSwain \& Gies~(\cite{mng05}).  

\section{Summary}
We report on a search for CBes from the \ldri.  A total of 192 (12 known CBes) 
objects were identified as CBes with prominent He~I~$\lambda4387$, 
He~I~$\lambda4471$, and Mg~II~$\lambda4481$ absorption lines, as well as 
H$\beta$~$\lambda4861$ and H$\alpha$~$\lambda6563$ emission lines.  These 
candidates significantly increased current CBe sample of about 8\%.  Most of 
the CBe candidates are distributed at the Galactic Anti-Center due to the LAMOST 
observing strategy.  Only two CBe candidates, L032 and L056, were found to be 
the memberships of the star clusters with ages of 15.8 and 398~Myr, respectively.

\acknowledgement{This research was supported by '973 Program' 2014 CB845702, 
the Strategic Priority Research Program ``The Emergence of Cosmological 
Structures'' of the Chinese Academy of Sciences, Grant No. XDB09000000 \& 
XDB09010100, the National Science Foundation of China (NSFC) under grants 
11173044 (PI:Hou), NSFC 14ZR1446900 (PI:Zhong), and NSFC 11390373 (PI:Shao).  
Guoshoujing Telescope (the Large Sky Area Multi-Object Fiber Spectroscopic 
Telescope LAMOST) is a National Major Scientific Project built by the Chinese 
Academy of Sciences. Funding for the project has been provided by the National 
Development and Reform Commission. LAMOST is operated and managed by the 
National Astronomical Observatories, Chinese Academy of Sciences.  This work 
is also supported in part by the Ministry of Science and Technology of Taiwan 
under grants NSC 103-2917-I-564-004 (Yu, P.-C.).}

\begin{table}
\begin{center}
\caption[]{Empirical $EW_{\lambda}$ of Known CBes in LAMOST}\label{ew}
 \begin{tabular}{cc}
  \hline\noalign{\smallskip}
Line &  $EW_{\lambda}$~(\AA)    \\
  \hline\noalign{\smallskip}
He~I~${\lambda4387}$   &  $0.387\pm0.185$ \\ 
He~I~${\lambda4471}$   &  $0.663\pm0.265$ \\ 
Mg~II~${\lambda4481}$  &  $0.291\pm0.141$ \\ 
H$\beta$~${\lambda4861}$  &  $<0.33$ \\ 
H$\alpha$~${\lambda6563}$  &  $<0.50$ \\ 
  \noalign{\smallskip}\hline
\end{tabular}
\end{center}
\end{table}

\begin{table}
\begin{center}
\caption[]{23 Known CBes Observed in the \ldri}\label{know}
 \begin{tabular}{cccrrrrrrc}
  \hline\noalign{\smallskip}
ID & DR1 & Designation & $J$ & $\Delta{J}$ & $H$ & $\Delta{H}$ & $K_{s}$ & $\Delta{K_{s}}$  & Remark \\
 & index & 2MASS & mag & mag & mag & mag & mag & mag &  \\
  \hline\noalign{\smallskip}
K01	&	{558648}	&	J063259.37+045622.5	&	9.158		&	0.027	&	9.158		&	0.036	&	9.087		&	0.032	&	detected\\
K02	&	{356675}	&	J035358.25+465351.8	&	9.287		&	0.022	&	9.049		&	0.028	&	8.826		&	0.023	&	detected\\
K03	&	{1542080}	&	J032910.40+312159.2	&	9.368		&	0.030	&	7.987		&	0.031	&	7.173		&	0.023	&	variant\\
K04	&	{555829}	&	J063337.49+044847.0	&	9.395		&	0.024	&	8.946		&	0.023	&	8.644		&	0.023	&	poor calibrated\\
K05	&	{567874}	&	J052314.90+374253.6	&	9.644		&	0.020	&	9.582		&	0.015	&	9.475		&	0.015	&	detected\\
K06	&	{500325}	&	J051502.46+364155.0	&	9.971		&	0.020	&	9.828		&	0.019	&	9.737		&	0.018	&	low $SNR$\\
K07	&	{436313}	&	J035447.92+445619.6	&	10.318	&	0.022	&	10.218	&	0.030	&	10.000	&	0.023	&	detected\\
K08	&	{1752150}	&	J055554.66+284706.3	&	10.966	&	0.036	&	10.914	&	0.033	&	10.872	&	0.027	&	detected\\
K09	&	{589106}	&	J063129.76+045449.1	&	11.449	&	0.024	&	10.887	&	0.025	&	9.690		&	0.021	&	low $SNR$\\
K10	&	{588035}	&	J063241.74+045338.4	&	11.927	&	0.021	&	11.822	&	0.025	&	11.664	&	0.024	&	missed\\
K11	&	{1745288}	&	J060559.66+280247.7	&	12.023	&	0.021	&	11.341	&	0.020	&	11.208	&	0.018	&	poor calibrated\\
K12	&	{510383}	&	J044927.22+450443.8	&	12.284	&	0.020	&	12.083	&	0.021	&	11.873	&	0.018	&	low $SNR$\\
K13	&	{1556719}	&	J051427.40+324756.8	&	13.802	&	0.030&	13.177	&	0.025	&	12.906	&	0.031&	poor calibrated\\   \hline
K14	&	{1718553}	&	J051214.46+411300.8	&	6.373		&	0.024&	5.896		&	0.033	&	5.621		&	0.016	&	poor calibrated\\
K15	&	{587291}	&	J063354.40+043935.2	&	6.996		&	0.020	&	6.945		&	0.040	&	6.866		&	0.023&	detected\\
K16	&	{1768331}	&	J070534.82+142831.7	&	7.156		&	0.020&	7.220		&	0.024	&	7.238		&	0.024&	detected\\
K17	&	{589139}	&	J063259.01+054756.6	&	7.640		&	0.023	&	7.390		&	0.049	&	6.966		&	0.020	&	low $SNR$\\
K18	&	{638021}	&	J151811.89+313849.2	&	7.907		&	0.020	&	7.713		&	0.031	&	7.706		&	0.016&	detected\\
K19	&	{1749404}	&	J054853.75+290801.7	&	7.959		&	0.024&	8.015		&	0.047	&	8.035		&	0.029&	detected\\
K20	&	{1496502}	&	J075704.21+025655.6	&	8.020		&	0.034	&	7.917		&	0.061	&	7.806		&	0.024	&	detected\\
K21	&	{1620649}	&	J062404.17+252508.1	&	8.042		&	0.024	&	7.916		&	0.016	&	7.752		&	0.018&	low $SNR$\\
K22	&	{321794}	&	J044440.68+503202.1	&	8.123		&	0.020	&	7.956		&	0.023	&	7.865		&	0.020&	detected\\
K23	&	{447292}	&	J065513.76+052554.4	&	8.293		&	0.024	&	8.425		&	0.047	&	8.440		&	0.021&	detected\\
\noalign{\smallskip}\hline
\end{tabular}
\tablecomments{850pt}{The K14--K23 are matched from the 2MASS point source 
catalog with radii larger than 5\arcsec.}
\end{center}
\end{table}


\begin{table}
\begin{center}
\caption[]{CBe Candidates}\label{can}
 \begin{tabular}{ccrrrrrrc}
  \hline\noalign{\smallskip}
ID &  Designation & $J$ & $\Delta{J}$ & $H$ & $\Delta{H}$ & $K_{s}$ & $\Delta{K_{s}}$  & SIMBAD \\
 &   & mag & mag & mag & mag & mag & mag &  \\
  \hline\noalign{\smallskip}
L001	&	 J043131.26+475750.7	&	9.062	&	0.027	&	8.721	&	0.015	&	8.435	&	0.019	&	EM* MWC 474, Em* \\
L002	&	 J043953.63+540146.1	&	9.102	&	0.021	&	8.867	&	0.015	&	8.598	&	0.014	&	BD+47 1000, Em* \\
L003	&	 J062753.84+003329.1	&	9.145	&	0.029	&	9.010	&	0.023	&	8.814	&	0.021	&	HD 291668 \\
L004	&	 J054520.88+290928.1	&	9.164	&	0.022	&	9.165	&	0.021	&	9.150	&	0.017	&	HD 247042 \\
L005	&	 J052948.83+373100.0	&	9.176	&	0.024	&	9.098	&	0.032	&	9.018	&	0.022	&	BD+37 1207, Em* \\
L006	&	 J050543.34+353110.7	&	9.181	&	0.023	&	9.068	&	0.031	&	8.861	&	0.023	&	HD 280498, Em* \\
L007	&	 J063131.81+053051.7	&	9.416	&	0.023	&	9.285	&	0.022	&	9.057	&	0.021	&	HD 258983, Em* \\
L008	&	 J044324.22+542816.5	&	9.471	&	0.021	&	9.119	&	0.015	&	8.891	&	0.021	&	TYC 3737-1292-1 \\
L009	&	 J054538.09+185753.7	&	9.508	&	0.021	&	9.500	&	0.024	&	9.484	&	0.022	&	HD 247221, Em* \\
L010	&	 J064433.60+045757.7	&	9.669	&	0.023	&	9.532	&	0.021	&	9.264	&	0.019	&	HD 263072, Em* \\
...&...&...&...&...&...&...&...&...\\
L032$^{1}$	&	 J065029.43+063621.0	&	10.453	&	0.026	&	10.404	&	0.023	&	10.328	&	0.023	&	TYC 160-841-1	\\
...&...&...&...&...&...&...&...&...\\
L056$^{2}$	&	 J051841.29+374030.0	&	10.975	&	0.019	&	10.966	&	0.028	&	10.946	&	0.026	&	HD 280870, Em*	\\
...&...&...&...&...&...&...&...&...\\
L171	&	 J053549.34+271917.2	&	14.988	&	0.036	&	14.574	&	0.049	&	14.451	&	0.063	&	new	\\
L172	&	 J003135.88+434905.3	&	15.252	&	0.044	&	15.185	&	0.087	&	15.111	&	0.123	&	V* HQ And, Nova	\\
L173	&	 J065742.53+175352.4	&	15.302	&	0.067	&	15.158	&	0.114	&	15.197	&	0.140	&	new	\\
L174	&	 J004339.36+411008.6	&	15.616	&	0.072	&	15.395	&	0.107	&	15.343	&	0.166	&	[HIB95] 29-13	\\
L175	&	 J003720.64+401637.6	&	15.745	&	0.068	&	15.486	&	0.126	&	15.374	&	0.196	&	new	\\
L176	&	 J004510.03+413657.6	&	15.766	&	0.070	&	15.636	&	0.122	&	15.562	&	0.177	&	new	\\
L177	&	 J052416.11+331819.9	&	15.803	&	0.073	&	15.363	&	0.103	&	15.006	&	0.140	&	new	\\
L178	&	 J004623.13+413847.4	&	15.829	&	0.064	&	15.697	&	0.128	&	15.602	&	0.170	&	LGGS J004623.14+413847.5	\\
L179	&	 J052432.28+332654.3	&	15.916	&	0.086	&	15.364	&	0.085	&	15.145	&	0.170	&	new	\\
L180	&	 J013420.91+303039.6	&	15.989	&	0.066	&	15.887	&	0.147	&	15.847	&	0.209	&	LGGS J013420.95+303039.9	\\
\noalign{\smallskip}\hline
\end{tabular}
\tablecomments{800pt}{\\  1. a member of the open cluster FSR\,1025 
                      \\  2. a member of the open cluster Kronberger\,18}
\end{center}
\end{table}

\label{lastpage}


\begin{thebibliography}{99}
\bibitem[1988]{bes88} Bessell, M.~S., \& Brett, J.~M.\ 1988, \pasp, 100, 1134 
\bibitem[2015]{cho15} Chojnowski, S.~D., Whelan, D.~G., Wisniewski, J.~P., et al.\ 2015, \aj, 149,7
\bibitem[2012]{cui12} Cui, X.-Q., Zhao, Y.-H., Chu, Y.-Q., et al.\ 2012, \raa, 12, 1197 
\bibitem[2003]{cut03} Cutri, R.~M., Skrutskie, M.~F., van Dyk, S., et al.\ 2003, yCat, 2246, 0
\bibitem[2005]{dre05} Drew, J.~E., Greimel, R., Irwin, M.~J., et al.\ 2005, \mnras, 362, 753 
\bibitem[2011]{eis11} Eisenstein, D.~J., Weinberg, D.~H., Agol, E., et al.\ 2011, \aj, 142, 72
\bibitem[2013]{kha13} Kharchenko, N.~V., Piskunov, A.~E., Schilbach, E., R{\"o}ser, S., \& Scholz, R.-D.\ 2013, \aap, 558, A53 
\bibitem[1997]{knw97} Kohoutek, L., \& Wehmeyer, R.\ 1997, Astron. Abh. Hamburger Sternw., 11, 1
\bibitem[1980]{liu80} Liu, C.-P., Zhang, C.-S., \& Kimura, H.\ 1980, Acta Astronomica Sinica, 21, 354 
\bibitem[2012]{luo12} Luo, A.-L., Zhang, H.-T., Zhao, Y.-H., et al.\ 2012, \raa, 12, 1243 
\bibitem[2015]{luo15} Luo, A.-L., et al.\ 2015, \raa, in press (arXiv:1505.01570)
\bibitem[2012]{maj12} Majewski, S.~R.\ 2012, American Astronomical Society Meeting Abstracts \#219, 219, \#205.06 
\bibitem[2005]{mng05} McSwain, M.~V., \& Gies, D.~R.\ 2005, \apjs, 161, 118 
\bibitem[2011]{nei11} Neiner, C., de Batz, B., Cochard, F., et al.\ 2011, \aj, 142, 149 
\bibitem[2003]{por03} Porter, J.~M., \& Rivinius, T.\ 2003, \pasp, 115, 1153 
\bibitem[2015]{rad15} Raddi, R., Drew, J.~E., Steeghs, D., et al.\ 2015, \mnras, 446, 274 
\bibitem[1985]{rie85} Rieke, G.~H., Cutri, R.~M., Black, J.~H., et al.\ 1985, \apj, 290, 116 
\bibitem[2013]{riv13} Rivinius, T., Carciofi, A.~C., \& Martayan, C.\ 2013, \aapr, 21, 69 
\bibitem[2015]{yup15} Yu, P.~C., Lin, C.~C., Chen, W.~P., et al.\ 2015, \aj, 149, 43 
\bibitem[2005]{zha05} Zhang, P., Chen, P.~S., \& Yang, H.~T.\ 2005, \na, 10, 325
\bibitem[2012]{zha12} Zhao, G., Zhao, Y.-H., Chu, Y.-Q., Jing, Y.-P., \& Deng, L.-C.\ 2012, \raa, 12, 723
\end{thebibliography}
\end{document}